\documentstyle[aps,prb,multicol,psfig]{revtex}
\begin{document}
\draft
\title{Critical behavior of thermopower and conductivity
at the metal-insulator transition in high-mobility Si-MOSFET's}

\author{R.\ Fletcher$^{a}$,  V.\ M.\ Pudalov$^b$
A.\ D.\ B.\ Radcliffe$^a$ and C.\ Possanzini$^c$}
\address{$^a$ Physics Department, Queen's University, Kingston,
Ontario, Canada, K7L 3N6.}
\address{$^b$ P.\ N.\ Lebedev Physics
Institute, 117924 Moscow, Russia.}
\address{$^c$ Physics Department, University of Nijmegen,
625 ED Nijmegen, The Netherlands.}
\date{\today}
\maketitle

\begin{abstract}

This letter reports thermopower and conductivity measurements
through the metal-insulator transition for 2-dimensional
electron gases in high mobility Si-MOSFET's.
At low temperatures both thermopower and conductivity
show critical behavior as a function of electron density
which is very similar to that expected for an Anderson transition.
In particular, when approaching the critical
density from the metallic side the diffusion thermopower appears
to diverge and the conductivity vanishes. On the insulating side
the thermopower shows an upturn with decreasing temperature.
\end{abstract}

\begin{multicols}{2}
Scaling theory of non-interacting, disordered, electron gases
predicts that no metal-insulator transition (MIT) occurs in 2
dimensions \cite{scaling,scalingreview,interact_scaling}
as temperature $T \rightarrow 0$. Nevertheless, what appears to be
a MIT has been observed (at finite, though low $T$), first in
$n-$Si-MOSFET's \cite{MOSFETrefs} and more recently in many other
2-dimensional (2D) hole and electron gases \cite{otherMIT}. In
the particular case of Si-MOSFET's, the transition is most clearly
visible in high-mobility samples, roughly $\mu \geq 1$\,m$^2$/V\,s.
As the density, $n$, is varied, there is a particular value,
$n_0$, above or below which the resistivity $\rho$ shows metallic or
insulating temperature dependence respectively. For the
present purposes we will use as a working definition that negative $d\rho/dT$
indicates an \lq insulator', and positive $d \rho/dT$ at the lowest
temperatures we can reach corresponds to a \lq metal'
(possible deviations from this definition and the consequences will be
mentioned later). At $n$ not too close to $n_0$, metallic behaviour
is visible over a wide range of $T$, roughly
$T < 0.5 E_F/k_B$ where $E_F$ is the Fermi energy.
The decrease of $\rho$ in the metallic state for high mobility
samples is typically two orders of magnitude larger than can be
accounted for by electron-phonon scattering.

Most previous work on these systems has focused on $\rho$,
though measurements of the compressibility \cite{compressibility}
have also appeared recently. The present paper presents
experimental data on the low temperature thermopower, $S$, and
conductivity, $\sigma =1/\rho$, both
of which are found to exhibit critical behavior around $n_0$.
Earlier, a scaling behaviour was described \cite{MOSFETrefs}
{\em for the temperature dependence of $\rho(T)$} over a temperature
range $\sim (0.05 - 0.3) E_F/k_B$.
In contrast, we report a different
type of critical behavior for $\sigma$. When we extrapolate our
data on $\sigma$, typically taken over the range $0.3-4.2$\,K, to the
$T\rightarrow 0$ limit, we find
{\em a power-law critical behaviour} as a function of $(n/n_0-1)$
on the metallic side. In addition, at our lowest temperature
of around 0.3\,K where diffusion thermopower dominates,
{\em $S/T$ appears to diverge when approaching $n_0$}.
At $n_0$ there is an abrupt change in behavior of $S/T$, with lower
densities showing an upturn in $S$ as $T$ is decreased.
Similar characteristics have long been predicted for an
Anderson MIT in 3D \cite{Mott,Castellani,Enderby,Villagonzalo}
but such a transition should not occur in 2D.

The main sample used for the present $\rho$ and $S$ measurements
(Sample 1) is the same as that described in a previous paper
\cite{paper1} and the techniques used to  measure $S$ can also
be found there.
This sample has $n_0 = 1.01 \times 10^{15}$\,m$^{-2}$ (as defined as above)
and a peak mobility $\mu = 1.75$\,m$^2$/Vs at $T = 1.1$\,K.
$S$ and $\rho$ have been measured as a function of $T$, down to about
0.3\,K, at many different values of $n$. We have also analyzed
independent $\rho(T,n)$ data for two other samples over the same range
of $T$, Sample 2 from the same wafer with
$n_0 = 0.99 \times 10^{15}$\,m$^{-2}$, and Sample 3
\cite{Pudalov} with peak $\mu = 3.6$\,m$^2$/Vs and
$n_0 = 0.956 \times 10^{15}$\,m$^{-2}$.
The major experimental problem was that of measuring
thermoelectric voltages with the sample in the insulating state. For
this purpose an amplifier with input bias current $< 1$\,pA and input
impedance  $>10^{12}\Omega$ was used. With some averaging it
had a resolution of 0.1\,$\mu$V
for source impedances of less than a few hundred k$\Omega$ rising to about
1\,$\mu$V at 10-20\,M$\Omega$, roughly the highest sample impedance in these
measurements. With the sample in the metallic state, a Keithley 182 digital
voltmeter usually gave the best compromise of input bias current, input
impedance and noise. All connections to the sample had isolation
resistance $ >50\,$G$\Omega$ and all leads were well shielded
and filtered against rf interference.

In the metallic region $n$ is a linear function of gate voltage and
it is believed to follow approximately the same dependence in the
insulating region \cite{hall}, at least close to $n_0$.
The results on the temperature dependence of $\rho$ on both the
insulating and metallic side are not shown
but are very similar to those seen in previous
work \cite{MOSFETrefs,Pudalov,VRHresistivity}.
In the metallic regime we have fitted our data on $\rho$, typically
over the range 0.3\,K to 4.2\,K, to the equation
\cite{Pudalov}
\begin{equation}
  \rho = \rho_0 + \rho_1 \exp \left(-(T_0/T)^p\right)
\end{equation}
where $\rho_0$, $\rho_1$, $T_0$ and $p$ are fitting constants,
in order to evaluate $\rho_0$.
\begin{figure}
\centerline{\psfig{figure=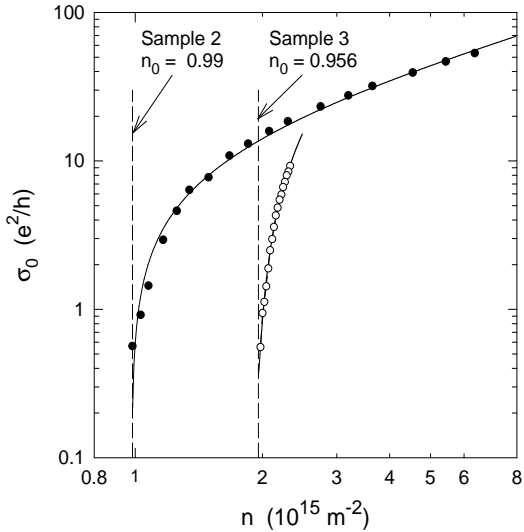,width=2.8in,height=2.9in}}
\begin{center}
\begin{minipage}{3in}
\caption{ Density dependence of the conductivity
in the  $T\rightarrow 0$ limit, of Samples 2
(solid symbols) and 3 (open symbols).
For sample 3 the data have been shifted by
adding $1 \times 10^{15}$\,m$^{-2}$ to $n$.}
\end{minipage}
\end{center}
\label{sigma}
\end{figure}
\vspace{-0.2cm}
Figure~1 shows the results on $\sigma_0 = 1/\rho_0$ as a function of $n$.
All samples follow the critical behaviour
\begin{equation}
\label{nscalesigma}
\sigma_0 = \sigma_m + \sigma_s\left(\frac{n}{n_0} - 1 \right)^\nu.
\end{equation}
The solid lines are the best fits with the following parameters
(with $\sigma$ in units of $e^2/h$).
For sample 2,  $\sigma_m = 0.2\pm 0.3$, $\sigma_s = 13.6\pm 0.7$
and $\nu=0.83 \pm 0.03$. Sample 1 has an identical behaviour within
experimental error. The higher-mobility Sample 3 also follows the
same equation with $\sigma_m=0.36\pm 0.15$, $\sigma_s = 34\pm 5$ and
$\nu=1.39\pm 0.05$. These results suggest $\nu $ increases with
peak mobility but clearly more
data on a variety of samples are required. The values of $\sigma_m$
for Sample 1 and 2 are consistent with zero within experimental
uncertainty. For Sample 3, $\sigma_m$ may be finite.
However, if $n_0$  is allowed to decrease from  $0.956$
to about $0.925
\times 10^{15}$\,m$^{-2}$, a fit which is indistinguishable over the
range of the data can also be obtained with $\sigma_m =0 \pm 0.15$,
$\sigma_s = 32\pm 5$ and $\nu=1.48\pm 0.05$. A small discrepancy in $n_0$
could arise from the identification of the critical density for the MIT
with that density, $n_0$, where $d\rho/dT$ changes sign, a procedure which
has no firm physical foundation \cite{amp2000}.
The critical behavior described by Eq.~(\ref{nscalesigma}) with
$\sigma_m = 0$ is formally the same as that expected for a (continuous)
Anderson transition with a mobility edge at $n_0$, whereas a finite
$\sigma_m$ would correspond to a (discontinuous) Mott-Anderson
transition; neither transition should arise in a non-interacting 2D
gas \cite{scaling,scalingreview}. The inclusion of interactions along
with disorder is a much more complex and ongoing theoretical problem
(e.g. see Refs. \onlinecite{scalingreview,interact_scaling}
and references therein) and it is not yet clear if such
transitions become possible under these conditions.
Similar critical behaviour, usually with $\sigma_m$ consistent with
zero, has been seen in many 3D systems,
typically with values \cite{Villagonzalo} of $\nu$ in the
range $0.5-1.3$. There are only two previously
reported cases related to 2D. Hanein {\em et al.} \cite{Hanein}
have made a similar analysis to the one above
for a 2D hole gas in GaAs and found a linear relation between $\sigma_0$
and $n$, but with a finite $\sigma_m$.
Feng {\it et al.} \cite{Feng}  have also found scaling behaviour in
a Si-MOSFET but it appears to be unrelated to that seen here.
\begin{figure}
\centerline{\psfig{figure=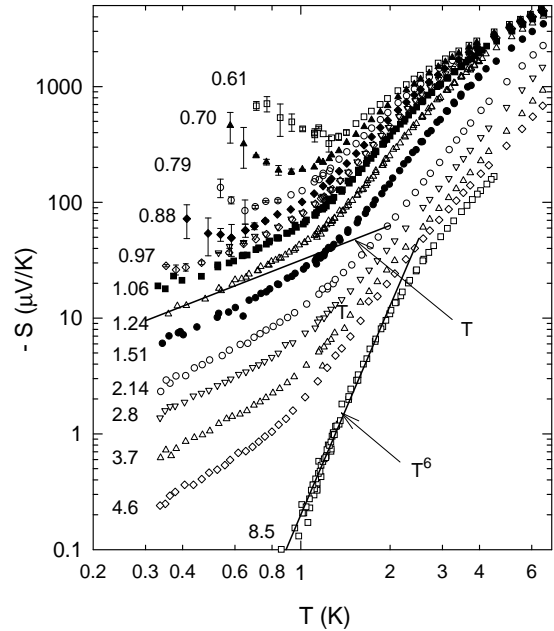,width=3.0in,height=3.5in}}
\begin{center}
\begin{minipage}{7.5cm}
\vspace{-0.2cm}
\caption{The thermopower $S$ for Sample 1 at various fixed
electron densities $n$ (in units of $10^{15}$\,m$^{-2}$).}
\end{minipage}
\end{center}
\label{TEP}
\end{figure}
\vspace{-0.5cm}

We now turn to the thermopower data. A selection of data on $S$ is
shown in Fig.~2. At
$n = 8.5\times 10^{15}$\,m$^{-2}$ diffusion thermopower, $S^d$,
is almost zero and one sees only phonon drag, $S^g$, which varies
approximately as $T^6$ at the lowest temperatures.\cite{paper1}
As $n$ decreases, $S$ begins to show two distinct regions with
different $T$ dependences. At $T > 1$\,K
there is a relatively rapid increase of $S$, roughly $\propto T^3$
which is that expected for $S^g$ at intermediate temperatures.\cite{paper1}
At $T < 1$\,K, $S$ has a much weaker, approximately linear $T$ dependence
indicative of $S^d$ becoming dominant; for $n < n_0$, this low-$T$ behaviour,
which is characteristic of ordinary metals, is replaced by an upturn in $S$.
Concentrating on the metallic region, the data at lowest $T$
are taken to give the best estimate of diffusion $S^d  = \alpha T$.
Fig.~3 shows that $\alpha$ as a function of $n$ appears
to diverge as $n \rightarrow n_0$. One would expect a divergence
when $E_F$ approaches a gap in the DOS but the present results
are inconsistent with this explanation because Hall data
\cite{hall} show that in the vicinity of $n_0$ the mobile carrier density
equals $n$ within $ 10\%$. However, Eq.~(\ref{nscalesigma}) also
implies a divergence of $S^d$. Thus, with the assumption of a
constant density of states (DOS) Eq.~(\ref{nscalesigma})
is consistent with
\begin{equation}
\label{escalesigma}
\sigma(E_F) = \sigma_m + \sigma_s \left(\frac{E_F}{E_c} - 1 \right)^\nu .
\end{equation}
Again, with $\sigma_m =0$ this
is formally equivalent to an Anderson transition with $E_c$ being
the mobility edge. The use of the Mott relation
$S^d = -(\pi^2 k_B^2 T/3e)(\partial \ln \sigma/\partial E)_{E_F}$
with Eq.~(\ref{escalesigma}) and taking $\sigma_m =0$ then gives
\cite{Castellani,Villagonzalo}
\begin{equation}
\label{metalsd}
 S^d = -\frac{\nu \pi^2 k_B^2 T}{3e(E_F- E_c)}~.
\end{equation}
This result is valid only if $(E_F - E_c)/k_BT \gg 1$;
in the opposite limit
$S^d$ tends to a constant \cite{Enderby,Villagonzalo}
($\sim 228\,\mu$V/K in 3D).
Numerical calculations\cite{Villagonzalo} show that the approximation
of Eq.~(\ref{metalsd}) gives a magnitude roughly a factor of 2 too
large when $(E_F - E_c)/k_BT \approx 2$ which, for our samples,
corresponds to $\Delta = (n-n_0)/n_0 \approx 0.11$ at $T = 0.4$\,K
(using the ideal DOS, $g_0$, with an effective mass of $0.19\,m_0$).
 \begin{figure}
\label{alpha}
\centerline{\psfig{figure=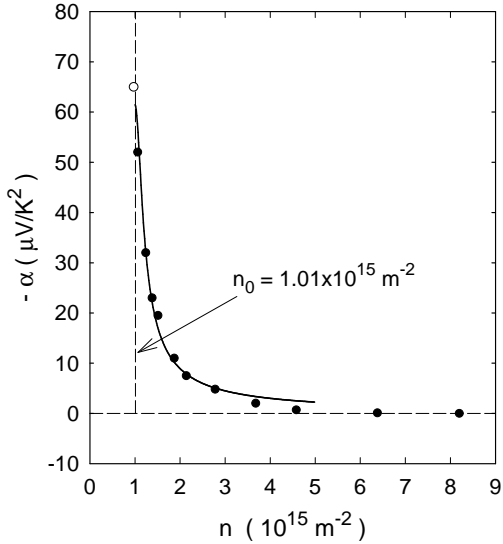,width=2.75in,height=3.0in}}
\begin{center}
\begin{minipage}{7.5cm}
\caption{
Density dependence of  $\alpha \approx S^d/T$ for Sample 1. The
single open symbol is for $n =0.97\times
10^{15}$\,m$^{-2}$ which is just below $n_0$ but $k_BT$ broadening
should make this indistinguishable from $n_0$. The line is the best fit
to Eq.~(\ref{scalesd}) for data with $n < 4\times 10^{15}$\,m$^{-2}$.}
\end{minipage}
\end{center}
\end{figure}
\vspace{-0.2cm}
To simulate this saturation we add $\Delta$ in the denominator
(but allow it to be a variable when determining the best
fit to the data) and, rewriting Eq.~(\ref{metalsd}) in terms
of $n$, we have
 \begin{equation}
 \alpha = S^d/T  =  -C /\sqrt{\Delta^2  + (\frac{n}{n_0} - 1)^2 }
\label{scalesd}
\end{equation}
where $C = \nu \pi^2 k_B^2 /(3eE_c)$ is a constant expected
to be about 32\,$\mu$V/K$^2$ for Sample 1, again using $g_0$.
If $\sigma_m$ is finite in Eq.~(\ref{escalesigma}), then the Mott
relation shows that it will contribute to the denominator of
Eq.~(\ref{metalsd}), also softening the divergence at $n=n_0$. However,
the experimental $\sigma_m$ is so small that this is negligible
compared to the finite-$T$ effect considered here.
The best fit of the data to Eq.~(\ref{scalesd}) gives
$\Delta = 0.15\pm 0.01$ and  $C = (9.5\pm 1.5)\,\mu$V/K$^2$
and is shown as the solid line in Fig.~3. (As with
$\sigma$, the fit can be improved if $n_0$ is slightly decreased).
$\Delta$ is consistent with that expected from the argument above,
but $C$ is too small by a factor of about 3. However, we emphasize that
we are comparing our results for a 2D system with a theoretical model
of an Anderson MIT valid for non-interacting electrons in 3D.
Some progress has been made on calculating $S^d$ with the inclusion
of weak interactions and disorder \cite{reizer}.
Corrections are found which are logarithmic in $T$ and difficult
to detect in thermopower; we are unable to explain the observed
strong density dependence in terms of the calculations.
We should mention that we can also represent the data over the same
range using the simple expression $\alpha = -56/n^{2.5}\,\mu$V/K$^2$,
with $n$ in units of $10^{15}$\,m$^{-2}$, but this has no obvious
physical explanation; in particular, it does not have the form that
we would expect
for $S^d$ approaching a band edge at $n = 0$, i.e. $S^d \propto 1/n$.

The data in the insulating regime also show a critical behaviour
qualitatively consistent with a mobility edge.  Thus the
observed upturn of $S^d$ is expected for activated conduction
across a mobility
gap with $(E_c-E_F) > k_BT$. Under these conditions the 3D Anderson
model predicts\cite{Mott} (see also the numerical calculations
in Ref. \cite{Villagonzalo})
\begin{equation}
\label{insulsd}
S^d = -\frac{k_B}{e}\left( A + \frac{E_c - E_F}{k_BT} \right)
\end{equation}
where $A$ is a constant of order unity.
For $n \leq 0.79\times 10^{15}$\,m$^{-2}$ the
observed minima in $S^d$ occur at $T_m$ consistent with
$(E_c-E_F)/k_B T_m \sim 2$. Eq.~(\ref{insulsd}) would
then imply that the values of $S^d$ at these points
should all have about the same magnitude.
However, the observed $S$ will have other contributions.
In particular there will be $S^g$ (see below) and also a
contribution to $S^d$ from variable range hopping
(VRH) through localized states.
(When two or more conduction mechanisms are present, the appropriate
$S^d$ are weighted by their contributions to $\sigma$).
The $T$ dependence of our $\rho$ data and other
previously published data \cite{VRHresistivity} in the insulating
region are consistent with Efros-Shklovskii VRH across a soft
Coulomb gap. For this mechanism one expects $S^d$
to be a constant given by\cite{B&CVRH,Zvyagin}
$S^d = -(k_B/e)(k_BT_0/C)(\partial \ln g(E)/\partial E)_{E_F}$
where $T_0$ can be obtained from the temperature dependence
of $\rho$, e.g. Ref.~\onlinecite{VRHresistivity},
$g(E)$ is the background DOS, and $C$ a constant $\approx 6$.
If we take  $(\partial \ln g(E)/\partial E)_{E_F} \sim 1/E_F$
(implying that $E_F$ may be in the tail of the DOS)
and again using $g_0$ to estimate $E_F$, we find that the
calculated  $S^d$ are typically a factor
of two smaller than the values of $S$ observed at $T_m$.
The argument is not significantly changed if Mott VRH is
assumed.\cite{B&CVRH,Zvyagin} In this case $S^d \propto T^{1/3}$
but the  magnitudes calculated for $S^d$ are similar.

As far as we are aware, the only previous work which attempted
to follow $S^d$ into the region of 2D electron localization was that
of Burns and Chaikin\cite{B&C} on thin films of Pd and PdAu. They found
an upturn of $S^d$ in the strong localization region but no divergence
at higher conductivities. The authors attributed their results to
the opening of a Mott-Hubbard gap. In 3D, Lauinger
and Baumann\cite{L&B} observed critical behaviour of $\sigma$ and
a divergence of $S^d$ for metallic AuSb films, but the magnitude of
the latter was 2 orders of magnitude smaller than seen here.
Other 3D experiments on SiP\cite{SiP} and NbSi\cite{NbSi}
saw no divergence on the metallic side.

For completeness, we make a few comments about $S$ at higher $T$
where $S^g$ is dominant. Little is known about the
behavior of $S^g$ near a MIT but it should be present
on the metallic side though its precise form is not
known\cite{EP2DS}. On the other hand, $S^g$ requires conservation
of crystal momentum for electron-phonon scattering so
that $S^g =0$ for conduction via VRH.\cite{Zvyagin,P&F} Thus,
$S^g$  should only exist on the insulator side when excitation
to delocalized states occurs. Our data show that at any fixed
$T \geq 2$\,K, $S$ rises as $n$ decreases but crosses $n_0$
smoothly, i.e., we no longer see divergent behavior of $S$
at $n_0$. These facts show that activated conduction must
be present for all densities $n < n_0$ that we have investigated,
even though $\rho$ data (both our own and those of
others\cite{VRHresistivity}) appear to follow the
Efros-Shklovskii VRH model.

In summary, the behavior of $\sigma$ and $S^d$ in the \lq metallic' and
\lq insulating' phases in Si-MOSFETs are surprisingly consistent with a
3D Anderson MIT, though such a transition is not expected to occur in a
2D electron gas. Nevertheless, it is important to remember that to
reliably identify the observed critical behaviour with a MIT requires
data in the zero $T$ limit. Although our analysis is based on an
extrapolation to zero $T$, the actual data extend only to 0.3K.
Thus we should be careful not to conclude that a mobility edge or a
MIT has necessarily been observed.
Even so, the present results provide new information on these systems
that further constrains any theoretical model proposed to explain the
MIT, whether such a transition be an apparent or real property as
$T \rightarrow 0$.

We acknowledge the support of the NSERC Canada, and from INTAS, RFBR,
NATO, Programs \lq Physics of nanostructures', \lq Statistical physics'
and \lq Integration'.

\end{multicols}

\end{document}